\documentclass[10pt,twocolumn,english,aps,pra,twocolum,superscriptaddress,bibnotes,amsmath,amssymb,floatfix]{revtex4-1}
\usepackage[colorlinks=true,citecolor=blue,linkcolor=magenta]{hyperref}
\usepackage[defaultcolor=red]{changes} 
\usepackage{soul}
\usepackage[utf8]{inputenc}
\usepackage[english]{babel}
\usepackage{amsmath,amsfonts,amssymb}
\usepackage[T1]{fontenc}
\usepackage{url}
\usepackage{graphicx}

\begin{document}

\title{Enhancement of vacuum-ultraviolet dispersive-wave emission\\using gas-filled tapered hollow-core fibers}

\author{Yinuo Zhao}
\thanks{These authors contributed equally to this work.}
\affiliation{State Key Laboratory of Ultra-intense laser Science and Technology, Shanghai Institute of Optics and Fine Mechanics (SIOM), Chinese Academy of Sciences (CAS), Shanghai 201800, China}
\affiliation{Center of Materials Science and Optoelectronics Engineering, University of Chinese Academy of Sciences, Beijing 100049, China}
\affiliation{Russell Centre for Advanced Lightwave Science, Shanghai Institute of Optics and Fine Mechanics and Hangzhou Institute of Optics and Fine Mechanics, Hangzhou, 311421, China}
\affiliation{Zhejiang Key Laboratory of Microstructured Specialty Optical Fiber, Hangzhou Institute of Optics and Fine Mechanics (HIOM), Hangzhou 311400, China}

\author{Donghan Liu}
\thanks{These authors contributed equally to this work.}
\affiliation{State Key Laboratory of Ultra-intense laser Science and Technology, Shanghai Institute of Optics and Fine Mechanics (SIOM), Chinese Academy of Sciences (CAS), Shanghai 201800, China}
\affiliation{Center of Materials Science and Optoelectronics Engineering, University of Chinese Academy of Sciences, Beijing 100049, China}
\affiliation{Russell Centre for Advanced Lightwave Science, Shanghai Institute of Optics and Fine Mechanics and Hangzhou Institute of Optics and Fine Mechanics, Hangzhou, 311421, China}
\affiliation{Zhejiang Key Laboratory of Microstructured Specialty Optical Fiber, Hangzhou Institute of Optics and Fine Mechanics (HIOM), Hangzhou 311400, China}

\author{Baoqi Shi}
\thanks{These authors contributed equally to this work.}
\affiliation{International Quantum Academy and Shenzhen Futian SUSTech Institute for Quantum Technology and Engineering, Shenzhen 518048, China}
\affiliation{Hefei National Laboratory, University of Science and Technology of China, Hefei 230088, China}

\author{Zhiyuan Huang}
\email[]{huangzhiyuan@siom.ac.cn}
\affiliation{State Key Laboratory of Ultra-intense laser Science and Technology, Shanghai Institute of Optics and Fine Mechanics (SIOM), Chinese Academy of Sciences (CAS), Shanghai 201800, China}
\affiliation{Russell Centre for Advanced Lightwave Science, Shanghai Institute of Optics and Fine Mechanics and Hangzhou Institute of Optics and Fine Mechanics, Hangzhou, 311421, China}
\affiliation{Zhejiang Key Laboratory of Microstructured Specialty Optical Fiber, Hangzhou Institute of Optics and Fine Mechanics (HIOM), Hangzhou 311400, China}

\author{Tiandao Chen}
\affiliation{State Key Laboratory of Ultra-intense laser Science and Technology, Shanghai Institute of Optics and Fine Mechanics (SIOM), Chinese Academy of Sciences (CAS), Shanghai 201800, China}

\author{Jinyu Pan}
\affiliation{State Key Laboratory of Ultra-intense laser Science and Technology, Shanghai Institute of Optics and Fine Mechanics (SIOM), Chinese Academy of Sciences (CAS), Shanghai 201800, China}

\author{Zhengzheng Liu}
\affiliation{State Key Laboratory of Ultra-intense laser Science and Technology, Shanghai Institute of Optics and Fine Mechanics (SIOM), Chinese Academy of Sciences (CAS), Shanghai 201800, China}

\author{Xinglin Zeng}
\affiliation{Russell Centre for Advanced Lightwave Science, Shanghai Institute of Optics and Fine Mechanics and Hangzhou Institute of Optics and Fine Mechanics, Hangzhou, 311421, China}
\affiliation{Zhejiang Key Laboratory of Microstructured Specialty Optical Fiber, Hangzhou Institute of Optics and Fine Mechanics (HIOM), Hangzhou 311400, China}

\author{Wenbin He}
\affiliation{Russell Centre for Advanced Lightwave Science, Shanghai Institute of Optics and Fine Mechanics and Hangzhou Institute of Optics and Fine Mechanics, Hangzhou, 311421, China}
\affiliation{Zhejiang Key Laboratory of Microstructured Specialty Optical Fiber, Hangzhou Institute of Optics and Fine Mechanics (HIOM), Hangzhou 311400, China}

\author{Jiapeng Huang}
\affiliation{Russell Centre for Advanced Lightwave Science, Shanghai Institute of Optics and Fine Mechanics and Hangzhou Institute of Optics and Fine Mechanics, Hangzhou, 311421, China}
\affiliation{Zhejiang Key Laboratory of Microstructured Specialty Optical Fiber, Hangzhou Institute of Optics and Fine Mechanics (HIOM), Hangzhou 311400, China}

\author{Jinxin Zhan}
\affiliation{Russell Centre for Advanced Lightwave Science, Shanghai Institute of Optics and Fine Mechanics and Hangzhou Institute of Optics and Fine Mechanics, Hangzhou, 311421, China}
\affiliation{Zhejiang Key Laboratory of Microstructured Specialty Optical Fiber, Hangzhou Institute of Optics and Fine Mechanics (HIOM), Hangzhou 311400, China}

\author{Xin Jiang}
\affiliation{Russell Centre for Advanced Lightwave Science, Shanghai Institute of Optics and Fine Mechanics and Hangzhou Institute of Optics and Fine Mechanics, Hangzhou, 311421, China}
\affiliation{Zhejiang Key Laboratory of Microstructured Specialty Optical Fiber, Hangzhou Institute of Optics and Fine Mechanics (HIOM), Hangzhou 311400, China}

\author{Yuxin Leng}
\affiliation{State Key Laboratory of Ultra-intense laser Science and Technology, Shanghai Institute of Optics and Fine Mechanics (SIOM), Chinese Academy of Sciences (CAS), Shanghai 201800, China}

\author{Junqiu Liu}
\email[]{liujq@iqasz.cn}
\affiliation{International Quantum Academy and Shenzhen Futian SUSTech Institute for Quantum Technology and Engineering, Shenzhen 518048, China}
\affiliation{Hefei National Laboratory, University of Science and Technology of China, Hefei 230088, China}

\author{Meng Pang}
\email[]{pangmeng@siom.ac.cn}
\affiliation{State Key Laboratory of Ultra-intense laser Science and Technology, Shanghai Institute of Optics and Fine Mechanics (SIOM), Chinese Academy of Sciences (CAS), Shanghai 201800, China}
\affiliation{Russell Centre for Advanced Lightwave Science, Shanghai Institute of Optics and Fine Mechanics and Hangzhou Institute of Optics and Fine Mechanics, Hangzhou, 311421, China}
\affiliation{Zhejiang Key Laboratory of Microstructured Specialty Optical Fiber, Hangzhou Institute of Optics and Fine Mechanics (HIOM), Hangzhou 311400, China}

\maketitle
\noindent\textbf{The recent breakthroughs in laser-driving $^{229}$Th nuclear transition have created an urgent demand for coherent vacuum-ultraviolet (VUV) sources delivering high spectral brightness at the critical 148.38 nm isomer energy. 
However, generating sufficient photon flux to overcome the low nuclear excitation probability remains a challenge for compact setups. 
While resonant dispersive wave emission in gas-filled hollow-core fibers offers a promising route, standard capillaries face a fundamental trade-off: 
maximizing input coupling requires large core diameters, whereas efficient nonlinear VUV conversion demands the high intensities using small cores.
Here, we resolve this conflict using a gas-filled tapered capillary fiber.
This architecture utilizes a longitudinally decreasing core diameter to combine a large input aperture with adiabatic field concentration, thereby continuously enhancing the nonlinear interaction. 
Experimentally, we demonstrate a widely tunable source (135--240 nm) that achieves a twofold efficiency enhancement specifically at the 148.38 nm wavelength compared to uniform geometries. 
By providing a scalable route to high-flux VUV generation, this work establishes a critical tabletop tool for advancing solid-state nuclear clocks and time-resolved spectroscopy.}

Ultrafast vacuum-ultraviolet (VUV) sources, offering high temporal resolution and broad spectral coverage, constitute a cornerstone for investigating fundamental light-matter interactions, ranging from chemical reaction dynamics to transient absorption spectroscopy \cite{Yang:15, Maiuri:19}.
In particular, the resonant excitation of the $^{229}\text{Th}$ nuclear clock transition has recently garnered widespread attention, promising a new paradigm for precision metrology \cite{VonderWense:16, Seiferle:19, Higgins:25} and fundamental physics tests \cite{Fuchs:25}.
Recent breakthroughs have successfully driven this nuclear transition using VUV laser sources, specifically targeting the isomer energy near 8.36 eV ($\sim$148.38 nm wavelength) across diverse platforms, ranging from trapped ions \cite{Yamaguchi:24} to solid-state crystals \cite{Kraemer:23, Tiedau:24, Elwell:24, Zhang:24, Hiraki:24} and thin films \cite{Zhang:24c}.

However, advancing nuclear spectroscopy and developing practical nuclear clocks impose stringent requirements on VUV sources: 
they must deliver high photon flux to overcome the low nuclear excitation probability, while simultaneously offering precise tunability to address the exact transition frequency. 
Consequently, the development of compact, efficient, and tunable coherent VUV sources around this spectral window is of urgent scientific importance.

In the pursuit of a nuclear clock, the requirements for source temporal properties are twofold: while continuous-wave lasers are indispensable for the final narrow-linewidth interrogation of the clock transition \cite{Xiao:26, Xiao:26b}, ultrafast pulsed sources are equally critical, serving as the foundation for VUV frequency combs---the ``clockwork'' required to link optical frequencies to electronic counters.
Currently, such coherent pulsed VUV radiation is primarily generated via free-electron lasers (FELs) \cite{Ayvazyan:02,Chang:18} or high-harmonic generation (HHG) \cite{Li:20,Yoshikawa:17,Zeng:24}.
While FELs offer unparalleled brightness and tunability, their accessibility is restricted to large-scale user facilities.
Conversely, tabletop HHG systems provide ultrashort pulse durations but typically suffer from low conversion efficiencies, necessitating complex high-energy driving lasers to achieve usable VUV fluxes.
To bridge the gap between these extremes, resonant dispersive wave (DW) emission in gas-filled hollow-core fibers (HCF) has emerged as a robust platform for generating tunable, few-cycle pulses with high brightness \cite{Russell:14,Travers:19,Pan:24,Chen:24,Chen:25}.
By balancing anomalous dispersion with self-phase modulation within a waveguide, soliton dynamics can drive efficient energy transfer to the DW in the VUV region.
The pioneering work by Travers \textit{et al}. has demonstrated microjoule-level pulse generation tunable from the deep- to vacuum-ultraviolet in a 3-m-long gas-filled capillary \cite{Travers:19}. 
Subsequent studies have extended this capability across the visible and ultraviolet regimes \cite{Brahms:20,Yu:25,Sabbah:25}.

Despite facilitating key applications \cite{Zhang:24b,Jackson:25,Russell:25}, scaling the conversion efficiency---particularly at shorter VUV wavelengths---remains a critical challenge for compact systems \cite{Brahms:24}.
Standard constant-diameter capillaries face an inherent trade-off: smaller core diameters are required to achieve the high intensities necessary for VUV generation, but they fundamentally limit the incoupling efficiency and energy handling capacity.

\begin{figure}[t!]
\centering
\includegraphics{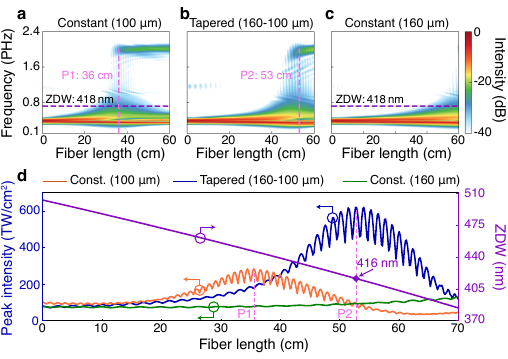}
\caption{
\textbf{Simulated spectral evolution with different capillary geometries}. 
Spectral evolution of the experimental ~10-fs pulse as a function of fiber length for three capillary configurations.
\textbf{a}. 
Constant 100-\textmu m-diameter capillary (0.6 m length) pumped with 100 \textmu J at 3.8 bar He. 
Purple dashed line indicates the ZDW at 418 nm. 
Pink dashed line marks the positions of maximum soliton compression at 36 cm (P1).
\textbf{b}.
Tapered capillary with core diameter decreasing linearly from 160 to 100 \textmu m (0.6 m length) pumped with 130 \textmu J at 3.25 bar He. 
Pink dashed line marks the positions of maximum soliton compression at 53 cm (P2).
\textbf{c}. 
Constant 160-\textmu m-diameter capillary (1 m length) pumped with 128 \textmu J at 1.48 bar He.
Purple dashed line indicates the ZDW at 418 nm. 
\textbf{d}. 
Peak intensity and ZDW evolution. 
Left axis: Simulated peak intensity (orange, blue, and green curves) versus fiber length for the three cases \textbf{a}, \textbf{b}, \textbf{c}. 
Right axis: ZDW evolution (purple line) for the tapered capillary. 
The intersection of P2 with the ZDW curve confirms that the ZDW at the DW emission point is 416 nm, nearly identical to that of the constant-core capillaries.
}
\label{Fig:1}
\end{figure}

Here, we overcome this limitation by demonstrating an efficient VUV source based on resonant DW emission in a compact, gas-filled tapered capillary fiber.
Unlike conventional geometries, our tapered waveguide features a longitudinally varying core size, starting with a large input diameter (160 \textmu m) to facilitate efficient coupling of high-energy pulses.
As the pulse propagates, the gradually decreasing core diameter adiabatically concentrates the optical field, continuously enhancing the waveguide nonlinearity.
This structural degree of freedom significantly elevates the peak intensity at the soliton fission point while maintaining favorable ionization dynamics.
Experimentally, we show that this intensity scaling enables highly efficient DW emission at the critical 148.38 nm band within a compact system footprint. 

\vspace{0.1cm}
\noindent\textbf{Theoretical analysis and simulation}.
The challenges associated with generating efficient VUV DW within compact geometries are governed by soliton scaling laws \cite{Russell:14,Travers:19,Heyl:16}. 
In standard capillaries, increasing the core diameter enables higher pulse-energy handling but at the expense of a longer zero-dispersion wavelength (ZDW) and reduced waveguide dispersion contribution.
Consequently, this red-shifted ZDW pushes the resonant DW emission frequency towards longer wavelengths, away from the targeted VUV spectral range.
To achieve phase-matching in the VUV region, the gas pressure must be relatively low to provide weak material dispersion that balances the waveguide dispersion \cite{Russell:14,Travers:19,Heyl:16}. 
While lower gas pressure and weaker dispersion facilitate energy scaling, they significantly extend the propagation distance required for pulse self-compression, thereby strictly limiting the compactness of the system \cite{Travers:24}.

\begin{figure}[t!]
\centering
\includegraphics{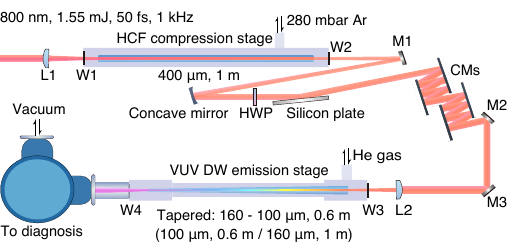}
\caption{
\textbf{Experimental setup}.
Schematic of the two-stage pulse compression and DW generation system. 
The input pulse is with 800 nm wavelength, 1.55 mJ energy, 50-fs duration  and 1 kHz repetition rate.
The hollow-core fiber (HCF) has 400 \textmu m core diameter and 1 m length. 
The tapered capillary of 0.6 m length has tapering core diameter from 160 \textmu m at the input to 100 \textmu m at the output.
In comparison, the constant-core capillaries have 100 \textmu m core diameter and 0.6 m length, or 160 \textmu m core diameter and 1 m length, respectively. 
L1--L2: plano-convex lenses; 
W1--W3: fused silica windows; 
W4: magnesium fluoride (MgF$_2$) window; 
HWP: half-wave plate; 
M1--M3: silver mirrors; 
CMs: chirped mirrors.
}
\label{Fig:2}
\end{figure}

We address this limitation by implementing a tapered capillary designed to accommodate higher input pulse energies while maintaining the nonlinearity required for efficient pulse compression. 
To investigate the pulse dynamics, we performed simulations based on the unidirectional pulse propagation equation \cite{Huang:18}, modeled using a fixed driving pulse duration of 10 fs at 800 nm wavelength. 
The results of these simulations are summarized in Fig.~\ref{Fig:1}.

First, as a benchmark, we simulate a constant-core (100 \textmu m diameter) capillary, as shown in Fig. \ref{Fig:1}a.
This geometry supports a maximum input energy of 100 \textmu J. 
To phase-match the DW emission at 148.38 nm, a He-gas pressure of 3.8 bar is required to position the waveguide ZDW at 418 nm. 
Under these conditions, the DW radiates at the maximum compression point ($\sim$36 cm), yielding a DW energy of $\sim$2.4 \textmu J with a conversion efficiency of $\sim$2.4\%.

Next, we examine the proposed tapered capillary, where the core diameter decreases linearly from 160 \textmu m to 100 \textmu m, as shown in Fig. \ref{Fig:1}b. 
The larger input core allows for an increased input pulse energy of 130 \textmu J. 
As the pulse propagates, the decreasing waveguide dimension enhances nonlinearity, accelerating the compression process. 
By optimizing the He-gas pressure to 3.25 bar, we ensure phase matching at 148.38 nm at the shifted maximum compression point ($\sim$53 cm). 
This configuration yields a strong DW emission with an output energy of 5.4 \textmu J and a conversion efficiency of $\sim$4.2\%.

\begin{figure}[t]
\centering
\includegraphics{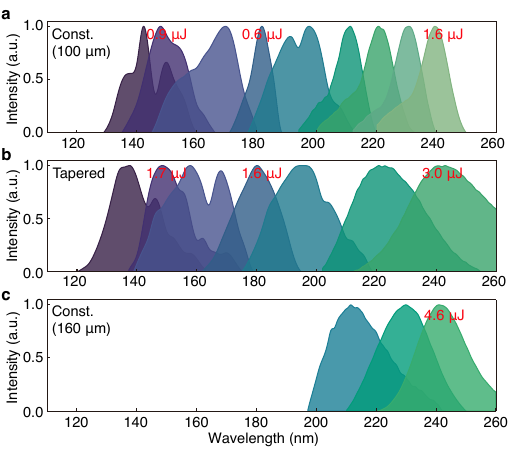}
\caption{
\textbf{Measured dispersive-wave spectra with different capillary geometries}. 
Evolution of the DW spectra as a function of tuning pressure for:
\textbf{a}.
Constant-core capillary of 100 \textmu m diameter. 
\textbf{b}.
Tapered-core capillary of 160 \textmu m to 100 \textmu m diameter. 
\textbf{c}.
Constant-core capillary of 160 \textmu m diameter.
The DW energies corresponding to central wavelengths of 148 nm, 180 nm, and 240 nm are explicitly highlighted in red.
}
\label{Fig:3}
\end{figure}

For comparison, we also simulate a constant-core capillary of 160 \textmu m diameter, as shown in Fig. \ref{Fig:1}c. 
While this core size supports higher energy, it requires a significantly lower pressure (1.48 bar) to achieve the same phase-matching conditions \cite{Yu:25,Travers:11}. 
This low pressure results in weak nonlinearity and dispersion, extending the pulse compression length well beyond the 1 m capillary length used in the simulation, thus precluding effective DW generation in a compact setup.

Consequently, the tapered capillary generates 148.38 nm DW pulses with 2.2-fold higher energy and 1.7-fold higher efficiency than that of the standard 100-\textmu m-diameter capillary. 
As shown in Fig. \ref{Fig:1}d, this enhancement arises from the combination of higher input energy capacity and reduced waveguide loss, both of which contribute to a significantly higher peak intensity at the moment of maximum compression.

\vspace{0.1cm}
\noindent\textbf{Experimental implementation}.
Guided by these numerical predictions, we experimentally validate the performance of the tapered capillary using the setup illustrated in Fig. \ref{Fig:2}. 
The system is seeded by a commercial Ti:Sapphire femtosecond laser (Spectra-Physics Solstice Ace), which delivers $\sim$50-fs pulses at 800 nm with a pulse energy of 1.55 mJ and a repetition rate of 1 kHz.
In the first stage, the laser output is spectrally broadened via self-phase modulation in a 1-m-long, Ar-filled capillary (400 \textmu m core diameter) and subsequently compressed to $\sim$10 fs using chirped mirrors.
These compressed 10-fs pulses are then coupled into the second-stage capillary for DW generation, allowing for direct comparison between the tapered and constant-core geometries.
The resulting output spectrum is characterized using a vacuum monochromator (McPherson 234/302) alongside a UV-NIR spectrometer (Ocean Optics Maya2000-Pro). 
Finally, the energy of the generated DW pulse is measured after isolation with a narrow-band optical filter \cite{Yu:25}.

\begin{figure}[t!]
\centering
\includegraphics{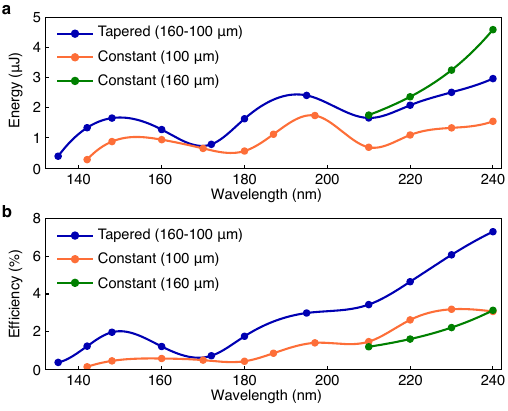}
\caption{
\textbf{Performance comparison}. 
\textbf{a}. 
Measured DW output energy as a function of wavelength. 
Blue curve: tapered-core capillary of 160 \textmu m to 100 \textmu m diameter. 
Orange curve: constant-core capillary of 100 \textmu m diameter. 
Green curve: constant-core capillary of 160 \textmu m diameter.
\textbf{b}. 
Measured DW conversion efficiency as a function of wavelength. 
}
\label{Fig:4}
\end{figure}

\vspace{0.1cm}
\noindent\textbf{Results}.
We evaluate three capillary geometries in the second stage, utilizing parameters consistent with the simulations. 
Figure \ref{Fig:3} summarizes the measured DW spectra for each case.
For the constant-core capillary (100~\textmu m diameter, 0.6 m length), varying the He-gas pressure from 1.58 to 9.73 bar allowed for spectral tuning from $\sim$142 nm to $\sim$240 nm (Fig. \ref{Fig:3}a) \cite{Yu:25,Travers:11}. 
To maintain appropriate soliton orders across this range, the driving pulse energy is adjusted between 46 and 202 \textmu J \cite{Russell:25}.
Similarly, the tapered capillary (160 \textmu m diameter at the input to 100 \textmu m diameter at the output, 0.6 m length) demonstrate a broad tuning range from 135 to 240 nm by adjusting the pressure from 2.37 to 7.75 bar (Fig. \ref{Fig:3}b).
In contrast, the large-core capillary (160 \textmu m diameter, 1 m length) is limited to generating DW pulses at wavelengths longer than 210 nm (Fig. \ref{Fig:3}c). 
Accessing wavelengths shorter than 210 nm in this geometry requires pressures below 1.85 bar; 
however, such low pressures reduce waveguide nonlinearity, extending the required compression length beyond the physical fiber length and preventing effective DW generation.
Consequently, the constant 160-\textmu m-diameter geometry is effectively precluded from accessing the VUV spectral region.

\begin{figure}[t!]
\centering
\includegraphics{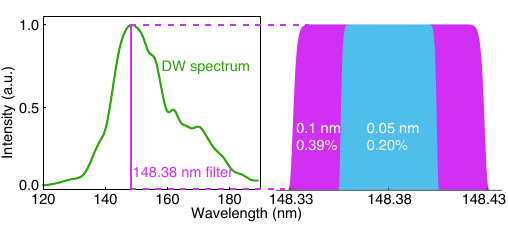}
\caption{
\textbf{Spectral power analysis at 148.38 nm wavelength}. 
Measured and normalized DW spectrum centered near the 148.38 nm $^{229}$Th transition. 
Shaded regions indicate the spectral energy retained after applying super-Gaussian filters centered at 148.38 nm with bandwidths of 0.05 nm (blue shading) and 0.1 nm (purple shading). 
These filtered bandwidths correspond to retained energy fractions of 0.20\% and 0.39\%, respectively.
}
\label{Fig:5}
\end{figure}

The performance metrics (output energy and conversion efficiency) are compared in Fig. \ref{Fig:4}. 
The tapered geometry (blue curve) consistently outperforms the constant 100-\textmu m-diameter capillary (orange curve) in terms of efficiency.
Specifically, at the target wavelength of 148.38 nm, the tapered capillary generates 1.7 \textmu J of DW energy from 84 \textmu J driving pulses, achieving a conversion efficiency of $\sim$2.0\%.
Conversely, the 100-\textmu m-diameter capillary requires a significantly higher driving energy of 194 \textmu J to generate just 0.9 \textmu J of DW output, resulting in a lower efficiency of $\sim$0.46\%.
Consequently, the tapered capillary experimentally demonstrates a 1.9-fold increase in output energy and a 4.3-fold enhancement in conversion efficiency compared to the standard capillary.

This disparity stems largely from coupling constraints. 
The tapered capillary benefits from a large entrance aperture, maintaining a robust coupling efficiency of $\sim$70\%. 
The 100-\textmu m-diameter capillary, however, suffered from limited coupling ($\sim$58\% at low energies), which dropped below 50\% at high driving energies due to ionization-induced defocusing at the fiber tip.
While the 160-\textmu m-diameter capillary (green curve) supports high driving energies and yields high output power at wavelengths $>210$ nm, the tapered capillary provides superior conversion efficiency across the tuning range of 140--240 nm.
We note that measured conversion efficiencies for all devices are lower than simulations, which we attribute to uncompensated capillary transmission losses and attenuation from optical components in the diagnostic path.

These results highlight the potential of high-flux VUV generation for $^{229}$Th nuclear clock applications \cite{VonderWense:16,Tiedau:24, Zhang:24}. 
To quantify this capability, we analyzed the spectral power at the critical 148.38 nm wavelength within the narrow bandwidths (0.05 nm and 0.1 nm) required for resonant nuclear $^{229}$Th excitation. 
In the current system, applying theoretical super-Gaussian filters with bandwidths of 0.05 nm and 0.1 nm yields average powers of approximately 3.4 \textmu W and 6.7 \textmu W, respectively, as shown in Fig. \ref{Fig:5}.

In conclusion, we have demonstrated a robust strategy for generating high-efficiency ultrafast VUV radiation by exploiting the unique properties of gas-filled tapered capillary fibers.
By implementing a longitudinally decreasing core diameter, we have successfully reconciled the trade-off between energy handling capacity and nonlinear compression efficiency that limits standard constant-core geometries.
At the critical 148.38 nm wavelength, the tapered device yields 1.7 \textmu J pulse energy with a conversion efficiency of $\sim$2.0\%---representing a nearly twofold increase in energy and a fourfold increase in efficiency compared to a standard 100-\textmu m-diameter capillary.
Furthermore, by linearly scaling this performance to a megahertz-repetition-rate driver, we project that milliwatt-level narrow-band VUV power at 148.38 nm is achievable, providing a robust source for future nuclear spectroscopy and clocks.

\medskip
\begin{footnotesize}

\noindent \textbf{Acknowledgments}: 
M. P. acknowledged the funding from National Natural Science Foundation of China (W2541021), the Strategic Priority Research Program of the Chinese Academy of Science (XDB0650000), the Shanghai Science and Technology Plan Project Funding (23JC1410100), the Shanghai Municipal Science and Technology Major Project.
J. L. acknowledged the funding from Shenzhen Science and Technology Program (Grant No. RCJC20231211090042078), 
Shenzhen-Hong Kong Cooperation Zone for Technology and Innovation (HZQB-KCZYB2020050), 
and Quantum Science and Technology--National Science and Technology Major Project (Grant No. 2023ZD0301500).

\noindent \textbf{Conflict of interest}: 
The authors declare no conflicts of interest.

\noindent \textbf{Data Availability Statement}: 
The code and data used to produce the plots within this work will be released on the repository \texttt{Zenodo} upon publication.
\end{footnotesize}

\vspace{0.3cm}

\bibliography{bibliography}

\end{document}